\documentclass[10pt]{iopart}

\begin{document}

\title{On the validity of the 5-dimensional Birkhoff theorem: The tale of an
exceptional case}
\author{Zolt\'{a}n Keresztes$^{1\dag }$ , L\'{a}szl\'{o} \'{A}. Gergely$%
^{1,2\ddag }$ \\
$^{1}$ Departments of Theoretical and Experimental Physics, University of
Szeged, D\'{o}m t\'{e}r 9, Szeged 6720, Hungary\\
$^{2}$ Department of Applied Science, London South Bank University, 103
Borough Road, London SE1 OAA, UK \\
$^{\dag }$zkeresztes@titan.physx.u-szeged.hu; $^{\ddag }$%
gergely@physx.u-szeged.hu}

\begin{abstract}
The 5-dimensional (5d) Birkhoff theorem gives the class of 5d vacuum
space-times containing spatial hypersurfaces with cosmological symmetries.
This theorem is violated by the 5d vacuum Gergely-Maartens (GM) space-time,
which is not a representant of the above class, but contains the static
Einstein brane as embedded hypersurface. We prove that the 5d Birkhoff
theorem is still satisfied in a weaker sense: the GM space-time is related
to the degenerated horizon metric of certain black-hole space-times of the
allowed class. This result resembles the connection between the
Bertotti-Robinson space-time and the horizon region of the extremal
Reissner-Nordstrom space-time in general relativity.
\end{abstract}

\section{Introduction}

In the most simple brane-world models the brane to which standard model
fields are confined is embedded into a 5-dimensional (5d) space-time, in
which only gravity acts. The basic dynamical equation on the brane is the
effective Einstein equation \cite{SMS}, supplemented by the Codazzi and
twice-contracted Gauss equations \cite{Decomp}. For a general overview of
brane-worlds see \cite{MaartensLR}.

Such models admit black hole solutions with tidal charge on the brane \cite%
{tidalRN}. The tidal charge represents the effect of the Kaluza-Klein modes
of gravity from the extra dimension, however the 5d space-time in which such
a brane is embedded, is still unknown. Stars on a brane \cite{GeMa}-\cite%
{Ovalle} and gravitational collapse under spherical symmetry \cite{BGM}-\cite%
{BraneOppSny} were also studied in brane-worlds, yielding to striking
features like the production of radiation in a spherically symmetric
collapse \cite{GovenderDadhich} and the emergence of unconventional forms of
stellar matter leading to dark energy production below the horizon \cite%
{BraneOppSny}.

Cosmological brane-world models were also studied, like a G\"{o}del brane 
\cite{BT}, branes with Swiss-cheese type inhomogeneities \cite{SwisScheese}-%
\cite{AsymmSwisScheese}, but most important, brane-world models with
Friedmann branes \cite{BDEL}, \cite{Decomp}, among them the static Einstein
brane \cite{GM}.

The Einstein static universe containing a perfect fluid is widely known to
be unstable against spatially homogeneous and isotropic perturbations \cite%
{Eddington}. However recent systematic analysis using covariant techniques 
\cite{Barrow} has shown that it is neutrally stable against small
inhomogeneous vector and tensor perturbations and neutrally stable against
adiabatic scalar density inhomogeneities for the velocity of sound obeying $%
c_{s}^{2}>1/5$. The stability of Einstein universes was also considered in
alternative gravitational theories, like the Einstein brane in the DGP model 
\cite{Shtanov}, in f(R) gravity \cite{Boehmer} and in Loop Quantum Cosmology 
\cite{Parisi}.

The most general static vacuum 5d space-time with cosmological constant $%
\widetilde{\Lambda }=3\varepsilon \Gamma ^{2}/\widetilde{\kappa }^{2}$ ($%
\varepsilon $ carries the sign of $\widetilde{\Lambda }$ and $\widetilde{%
\kappa }^{2}$ is the gravitational constant in 5d), which contains a
Friedmann brane and has the symmetries of the brane in each point is \cite%
{BCG} (see also \cite{BraneBlackHole}): 
\begin{eqnarray}
d\widetilde{s}^{2}=&-&f\left( r;k,\varepsilon \right) dt^{2}+\frac{dr^{2}}{%
f\left( r;k,\varepsilon \right) }\nonumber \\
&+&r^{2}\left[ d\chi ^{2}+\mathcal{H}%
^{2}\left( \chi ;k\right) \left( d\theta ^{2}+\sin ^{2}\theta d\varphi
^{2}\right) \right]   ,  \label{Sch}
\end{eqnarray}%
with the metric functions 
\begin{equation}
f\left( r;k,\varepsilon \right) =k-\frac{2m}{r^{2}}-\frac{\varepsilon \Gamma
^{2}}{2}r^{2}  ,
\end{equation}%
and 
\begin{equation}
\mathcal{H}\left( \chi ;k\right) =\left\{ 
\begin{array}{c}
\sin \chi ~,\qquad k=1 \\ 
\quad \chi ~~~,\qquad k=0 \\ 
~\sinh \chi ~,\qquad k=-1%
\end{array}%
\right. ~.  \label{H}
\end{equation}%
Here both $\varepsilon $ and $k$ take any of the values $\left( 0,\pm
1\right) $. This result is frequently referred as the 5d Birkhoff theorem.

However an interesting exceptional case has been found in \cite{GM},
representing a family of vacuum solutions of the 5d Einstein equations with
cosmological constant $\widetilde{\Lambda }$ which contain an Einstein
brane. This 5d space-time is given for $y>0$ as 
\begin{equation}
\Gamma ^{2}d\widetilde{s}^{2}=-F^{2}\left( y;\varepsilon \right) d\tau
^{2}+dy^{2}+d\chi ^{2}+\mathcal{H}^{2}\left( \chi ;\varepsilon \right)
\left( d\theta ^{2}+\sin ^{2}\theta d\varphi ^{2}\right) ~.  \label{GM}
\end{equation}%
The boundary at $y=0$ is the static Einstein brane \cite{GM}, therefore a
particular case of the Friedmann branes for which the 5d Birkhoff theorem
refers. The metric functions are 
\begin{equation}
F\left( y;\varepsilon \right) =\left\{ 
\begin{array}{c}
A\cos \left( \sqrt{2}y\right) +B\sin \left( \sqrt{2}y\right) ~,\qquad
\varepsilon =1 \\ 
A+\sqrt{2}By~\qquad \qquad \qquad \qquad ,\qquad \varepsilon =0 \\ 
A\cosh \left( \sqrt{2}y\right) +B\sinh \left( \sqrt{2}y\right) ~,\qquad
\varepsilon =-1%
\end{array}%
\right.   ,  \label{F}
\end{equation}%
and $\mathcal{H}\left( \chi ;\varepsilon \right) $ defined as (\ref{H}),
with $\varepsilon $ in place of $k$. A homogeneous counterpart of the
Gergely-Maartens (GM) metric (\ref{F}) was also found \cite{HomBrane}. The
GM metric is well defined on the brane for any $A\neq 0$. According to the
Lanczos equation a vanishing $B$ would be incompatible with brane matter 
\cite{GM}. As any of the constants $A$ or $B$ can be absorbed into the
coordinate $\tau $, the GM\ metric represents a one parameter family of
solutions, which is \textit{not} a sub-case of the metrics (\ref{Sch}).

The higher dimensional Birkhoff theorem was formulated in an alternative way
in \cite{5DBirkhoff}, by enouncing the set of conditions under which the
higher-dimensional space-time is static. The GM metric does not obey these
conditions either as its metric coefficient $g_{\chi \chi }$ is a constant.
Therefore it stays outside the validity of the Theorem 1 of \cite{5DBirkhoff}%
.

As remarked in Section 4 of \cite{GM}, the proof presented in \cite{BCG}
leading to the metric (\ref{Sch}) cannot be applied when the metric function 
$B$ of \cite{BCG} (which is different from the parameter $B$ of the metric (%
\ref{GM})) is a constant. Then it is not possible to introduce $r=B^{1/3}$
as a new radial coordinate in order to obtain the class of metrics (\ref{Sch}%
). This suggests however that the GM metric may be related to a tiny layer $%
\left( r,r+dr\right) $\ of the space-times (\ref{Sch}). It is the purpose of
the present paper to prove this conjecture and re-establish the validity of
the 5d Birkhoff theorem in a weaker sense.

In Section 2 we will enlist arguments in favour of the claim that the GM
space-time is related to the \textit{horizon} regions of certain 5d black
hole metrics (\ref{Sch}). We also write up an approximate form of the black
hole metrics (\ref{Sch}), valid in the vicinity of the degenerated horizons
of (\ref{Sch}). We present the horizons of the various metrics in the class (%
\ref{Sch}) in Appendix A. Section 3 contains the technical derivation of the
coordinate transformation bringing the black hole horizon metric into the GM
space-time, done explicitly for various subcases of the parameters of the GM
metric. Section 4 contains discussions on the equivalence of the GM metric
and horizon metric, based on the analysis of the Killing algebras, presented
in Appendix B. Section 5 is the concluding remarks. In Appendix C we present
a related result from general relativity: the Bertotti-Robinson solution 
\cite{Bertotti}, \cite{Robinson} describing gravity in the presence of a
covariantly constant electromagnetic field corresponds to the horizon region
of the extremal Reissner-Nordstr\"{o}m black hole \cite{Brill}. We present
this both for didactical reasons, as this derivation is not well known, and
as a simpler analogy for the method we follow in Section 3.

\section{Black hole horizons in a 5d space-time with Friedmann brane boundary%
}

It is immediate to see from the $\left( \chi ,~\theta ,~\varphi \right) $
sector that any relation between the space-times (\ref{Sch}) and (\ref{GM})
may exist only for $k=\varepsilon $. In what follows, we will discuss only
such metrics from the class (\ref{Sch}).\footnote{%
As we show in Appendix A, only a subset of these metrics have horizons. In
the cases $km>0$ the metric (\ref{Sch}) is also known as 5d topological
black hole (TBH) for $\varepsilon =0$ or (anti) de Sitter TBH for $%
\varepsilon =1$ ($\varepsilon =-1$), see \cite{SW}.} As noted in \cite{GM},
the curvature scalar of the GM solution with $\varepsilon =-1$ agrees with
the curvature scalar\ of the 5d black hole metric (\ref{Sch}) with $%
k=\varepsilon =-1$, only when evaluated at the horizon (when $m=-1/4\Gamma
^{2}$). This is a serious indication that the GM solution is related to the
event horizon of certain 5d black hole metrics.

No such relation exists in the non-cosmological case ($\varepsilon =0$).
Then the scalars $\widetilde{R}_{abcd}\widetilde{R}^{abcd}$ and$\ \widetilde{%
C}_{abcd}\widetilde{C}^{abcd}$ vanish in the GM space-time, however in the
space-time (\ref{Sch}) they are $\widetilde{R}_{abcd}\widetilde{R}^{abcd}=%
\widetilde{C}_{abcd}\widetilde{C}^{abcd}=288m^{2}/r^{8}.$ This can vanish
for any finite value of $r$ only if $m=0$, but then the metric (\ref{Sch})
becomes ill-defined.

In order to establish more exactly the connection between the GM space-time
and the horizon regions of the SchwarzSchild - (anti) de Sitter metrics (\ref%
{Sch}) we enlist the loci of the horizons (given by $f=0$) for various $%
\varepsilon $ in the Tables \ref{table1}-\ref{table3} of Appendix A. We note
that there is no horizon in the case $\varepsilon =0$, so it is not
surprising that in this case no connection can be established with the
family of GM metrics.

There are two horizons only in the cases $\varepsilon =k=1$, $m>0$ or $%
\varepsilon =k=-1$, $m<0$. These horizons merge into one (degenerated)
horizon at $\Gamma r=1$ for $\varepsilon m=1/4\Gamma ^{2}$. The latter is
exactly the condition, under which for $\varepsilon =-1$ the curvature
scalar\ of the 5d black hole metric was shown to agree with the curvature
scalar of the GM metric. Therefore we expect to find a correspondence
between the GM metric and the \textit{degenerated horizon regions} of the 5d
black hole metrics.

For this we introduce the new coordinate $\rho =\Gamma r-1$, which is small
close to the degenerated horizon, positive above the horizon and $\rho \in
\left( -1,0\right) $ below the horizon. For small $\rho $ the metric
function $f$ has the approximate expression $f=-2\varepsilon \rho ^{2}$ and
by rescaling the time coordinate as $t\rightarrow 4\Gamma ^{2}t$ we obtain
the "horizon metric"%
\begin{equation}
\Gamma ^{2}d\widetilde{s}^{2}=\frac{\varepsilon }{2}\left( \rho ^{2}dt^{2}-%
\frac{d\rho ^{2}}{\rho ^{2}}\right) +d\chi ^{2}+\mathcal{H}^{2}\left( \chi
;\varepsilon \right) \left( d\theta ^{2}+\sin ^{2}\theta d\varphi
^{2}\right)   ,  \label{Horizons_sol}
\end{equation}%
describing for $\varepsilon =k=1$ (or $\varepsilon =k=-1$) the vicinity of
the horizon of the SchwarzSchild - de Sitter (or Schwarzschild - anti de
Sitter-like with $k=-1$) space-time. The time coordinate for $\varepsilon =1$
is $\rho $ and for $\varepsilon =-1$ is $t$. We give the Killing vectors and
Killing algebra of the horizon metric in Appendix B. As the horizon metric
solves the $5$-dimensional Einstein equations in the presence of a
cosmological constant $\widetilde{\Lambda }=3\varepsilon \Gamma ^{2}/%
\widetilde{\kappa }^{2}$, it can be extended towards non-small values of $%
\rho $ either.

\section{The relation between the GM and the horizon metrics}

In this section we prove that for $\varepsilon =\pm 1$ the GM space-time is
related to the degenerated horizon region (\ref{Horizons_sol}) of the 5d
black hole (\ref{Sch}) with the same $\varepsilon $ and $k=\varepsilon $. In
order to compare the degenerated horizon region (\ref{Horizons_sol}) of the
5d black hole metric (\ref{Sch}) with the GM metric, we absorb its parameter 
$A$ into $\tau $ and denote $B/A$ with $B$. Then we rewrite the metric
function $F\left( y;\varepsilon =\pm 1\right) $ as%
\begin{eqnarray}
F\left( y;\varepsilon \right) &=&\cos z+\beta \sin z~,\qquad \varepsilon
=\pm 1  ,  \nonumber \\
z\left( y;\varepsilon \right) &=&\sqrt{2}~i^{\left( 1-\varepsilon \right)
/2}~y  \nonumber \\
\beta \left( B;\varepsilon \right) &=&\left( -i\right) ^{\left(
1-\varepsilon \right) /2}B  \label{Fpm}
\end{eqnarray}

Next we try to identify a suitable coordinate transformation $\left( t,\rho
\right) \rightarrow \left( \tau ,y\right) $ of the horizon metric. In order
to enforce the correspondence with the GM space-time, the original
coordinates $t\left( \tau ,y\right) $, $\rho \left( \tau ,y\right) $ have to
obey the following differential equations
\label{diffe}
\begin{eqnarray}
\rho ^{2}\left( \frac{\partial t}{\partial \tau }\right) ^{2}-\frac{1}{\rho
^{2}}\left( \frac{\partial \rho }{\partial \tau }\right) ^{2}
&=&-2\varepsilon F^{2}\left( y;\varepsilon \right)   ,  \label{diffe1}
\\
\rho ^{2}\left( \frac{\partial t}{\partial \tau }\right) \left( \frac{%
\partial t}{\partial y}\right) -\frac{1}{\rho ^{2}}\left( \frac{\partial
\rho }{\partial \tau }\right) \left( \frac{\partial \rho }{\partial y}%
\right) &=&0  ,  \label{diffe2} \\
\rho ^{2}\left( \frac{\partial t}{\partial y}\right) ^{2}-\frac{1}{\rho ^{2}}%
\left( \frac{\partial \rho }{\partial y}\right) ^{2} &=&2\varepsilon   %
.  \label{diffe3}
\end{eqnarray}%
For separable solutions $t=t_{0}\left( \tau \right) t_{1}\left( y\right) $
and $\rho =\rho _{0}\left( \tau \right) \rho _{1}\left( y\right) $ the
system (\ref{diffe}) simplifies to 
\label{diffes}
\begin{eqnarray}
\left( \rho _{0}\dot{t}_{0}\right) ^{2}\left( \rho _{1}t_{1}\right)
^{2}-\left( \frac{\dot{\rho}_{0}}{\rho _{0}}\right) ^{2} &=&-2\varepsilon
F^{2}\left( y;\varepsilon \right)   ,  \label{diffea} \\
\left( \rho _{0}^{2}t_{0}\dot{t}_{0}\right) \left( \rho
_{1}^{2}t_{1}t_{1}^{\prime }\right) -\left( \frac{\dot{\rho}_{0}}{\rho _{0}}%
\right) \left( \frac{\rho _{1}^{\prime }}{\rho _{1}}\right) &=&0  ,
\label{diffeb} \\
\left( \rho _{0}t_{0}\right) ^{2}\left( \rho _{1}t_{1}^{\prime }\right)
^{2}-\left( \frac{\rho _{1}^{\prime }}{\rho _{1}}\right) ^{2}
&=&2\varepsilon   ,  \label{diffec}
\end{eqnarray}%
where a dot (a prime) denotes the derivative with respect to $\tau $ ($y$).
In the last equation only $\rho _{0}^{2}t_{0}^{2}$ depends on $\tau $,
therefore either (a) $t_{1}^{\prime }=0$, thus $t=t\left( \tau \right) $ or
(b) $\rho _{0}t_{0}$=const . We consider these cases separately:

(a) When $t=t\left( \tau \right) $ from (\ref{diffec}) 
\begin{equation}
\rho _{1}=C_{1}\exp \left( \pm \sqrt{-2\varepsilon }y\right)   ,
\end{equation}%
($C_{1}$ a constant). Eq. (\ref{diffeb}) gives $\rho _{0}\left( \tau \right)
=C_{0}$ (a constant). Substituting $\rho $ into (\ref{diffea}) finally we
get 
\begin{equation}
C_{0}C_{1}\exp \left( \pm \sqrt{-2\varepsilon }y\right) \left( \frac{dt}{%
d\tau }\right) =\sqrt{-2\varepsilon }F\left( y;\varepsilon \right)   .
\label{F2}
\end{equation}%
Thus $dt/d\tau $ must be another constant, say $C_{2}$ and%
\begin{equation}
F\left( y;\varepsilon \right) =\frac{C_{0}C_{1}C_{2}}{\sqrt{-2\varepsilon }}%
\exp \left( \pm \sqrt{-2\varepsilon }y\right)   .  \label{F2B}
\end{equation}%
For the particular values of the constants $C_{0}C_{1}C_{2}=\sqrt{2}$ and
for $\varepsilon =-1$ this becomes the metric function (\ref{Fpm}) for $%
B=\pm 1$: 
\begin{equation}
F\left( y;-1\right) =\exp \left( \pm \sqrt{2}y\right) =\cosh \sqrt{2}y\pm
\sinh \sqrt{2}y  .  \label{FmB1}
\end{equation}%
If we additionally choose $C_{2}=1$, then%
\begin{equation}
t=\tau ~, { \quad }\rho =\sqrt{2}\exp \left( \pm \sqrt{2}y\right)  { 
}  \label{coorda}
\end{equation}%
is a coordinate transformation from the degenerated horizon metric (\ref%
{Horizons_sol}) with $\varepsilon =-1$ into the GM space-time with $%
\varepsilon =-1$ and $B=\pm 1$.

(b) For $t_{1}^{\prime }\neq 0$ and $\rho _{0}t_{0}=C_{3}\neq 0$ (a
constant) Eq. (\ref{diffea}) becomes%
\begin{equation}
\frac{\dot{\rho}_{0}^{2}}{\rho _{0}^{2}}=\frac{-2\varepsilon F^{2}\left(
y;\varepsilon \right) }{\left( \rho _{1}^{2}t_{1}^{2}C_{3}^{2}-1\right) }%
  .  \label{aa}
\end{equation}%
This gives $\dot{\rho}_{0}/\rho _{0}=-D$ (a constant), thus%
\begin{equation}
\rho _{0}=C_{4}\exp \left( -D\tau \right) , { \quad }t_{0}=\frac{C_{3}}{%
C_{4}}\exp \left( D\tau \right)   ,
\end{equation}%
where $C_{4}$ is the integration constant. Eq. (\ref{aa}) also implies 
\begin{equation}
t_{1}^{2}=\frac{D^{2}-2\varepsilon F^{2}\left( y;\varepsilon \right) }{%
C_{3}^{2}D^{2}\rho _{1}^{2}}  ,  \label{t1mo}
\end{equation}%
and from $\rho _{0}t_{0}=C_{3}$ and Eq. (\ref{diffeb})%
\begin{equation}
\rho _{1}^{2}t_{1}t_{1}^{\prime }C_{3}^{2}+\frac{\rho _{1}^{\prime }}{\rho
_{1}}=0  .  \label{rho1}
\end{equation}%
The last two equations imply 
\begin{equation}
\frac{\rho _{1}^{\prime }}{\rho _{1}}=\frac{F^{\prime }\left( y;\varepsilon
\right) }{F\left( y;\varepsilon \right) }  ,
\end{equation}%
with solution%
\begin{equation}
\rho _{1}=GF\left( y;\varepsilon \right)   ,
\end{equation}%
where $G$ is an integration constant. Finally Eq. (\ref{diffec}) constraints 
$F\left( y;\varepsilon \right) $ as%
\begin{equation}
2\varepsilon F^{\prime }\left( y;\varepsilon \right) ^{2}+4F\left(
y;\varepsilon \right) ^{2}-2\varepsilon D^{2}=0  .  \label{Feq}
\end{equation}%
The metric function (\ref{Fpm}) solves this equation for $D^{2}=2\left(
B^{2}+\varepsilon \right) $. Thus for this value of $D$ the horizon metric
transforms into the GM metric under the coordinate transformation%
\begin{eqnarray}
\rho &=&\sqrt{2}\exp \left( -D\tau \right) F\left( y;\varepsilon \right) ~, 
\nonumber \\
t &=&\frac{\left( -i\right) ^{\left( 1-\varepsilon \right) /2}}{\sqrt{2}%
\left( B^{2}+\varepsilon \right) ^{1/2}}\exp \left( D\tau \right) \frac{\tan
z-\beta }{1+\beta \tan z}~.  \label{coordb}
\end{eqnarray}%
(We have set $C_{4}G=\sqrt{2}$.)

We note that the result derived in (a) only partially emerges from the limit 
$D\rightarrow 0\,$\ of the result derived in (b) specified for $\varepsilon
=-1$ and $B=\pm 1$ (thus $D=0$) in the following sense. First, for $D=0$ the
metric function (\ref{FmB1}) solves the differential equation (\ref{Feq}).
Second, the expression of $\rho $ from (\ref{coordb}) reduces to the
corresponding expression (\ref{coorda}), however the transformation from $t$
to $\tau $ differs in a shift $1/D\rightarrow \infty $: \ 

\section{Discussion}

The transformation (\ref{coordb}) admits the following three particular
cases:

(b1) Case $\varepsilon =1.$ Then the horizon coordinates $\left( t,~\rho
\right) $ are related by a real coordinate transformation to the GM
coordinates $\left( \tau ,~y\right) $:%
\begin{eqnarray}
\rho &=&D\exp \left( -D\tau \right) \cos \left( \alpha _{1}+\sqrt{2}y\right) 
  ,  \nonumber \\
t &=&\frac{1}{D}\exp \left( D\tau \right) \tan \left( \alpha _{1}+\sqrt{2}%
y\right)   ,  \label{coordt1}
\end{eqnarray}%
where we have denoted $B=-\tan \alpha _{1}$. We also note that this
transformation obeys $t^{2}\rho ^{2}<1$, thus the GM space-time only
partially covers the horizon space-time (\ref{Horizons_sol}).\footnote{%
This is similar to an other famous example from brane-worlds, which
establishes the equivalence of the static branes written in Gauss normal
coordinates and moving branes in 5d SchwarzSchild - anti de Sitter
space-time\ \cite{Mukohyama}.}

Thus the transformation (\ref{coordt1}) links the $t^{2}\rho ^{2}<1$ region
of the 5d black hole horizon metric with $\varepsilon =1$ to the GM metric
with $\varepsilon =1$ . In this region\footnote{%
The 5d black hole metric (\ref{Sch}) for $\varepsilon =k=1$ and $m>0$ is
static only between the horizons (which degenerate for $m=1/4\Gamma ^{2}$).
In other worlds the Killing vector $K_{\mathbf{7}}$ is time-like between the
horizons, space-like outside. Thus for degenerated horizons there is no
time-like Killing vector. However the approximate degenerate horizon metric (%
\ref{Horizons_sol}) acquires new symmetries, among which $K_{\mathbf{8}}$ is
time-like for $t^{2}\rho ^{2}<1$.} the horizon metric is static due to $K_{%
\mathbf{8}}$, as show in Appendix B.

(b2) For $\varepsilon =-1$ and $B^{2}>1$ (implying $\mathit{{sgn}\left(
D^{2}\right) =1}$) the coordinate transformation is%
\begin{eqnarray}
\rho &=&D\exp \left( -D\tau \right) \sinh \left( \alpha _{2}+\sqrt{2}%
y\right)   ,  \nonumber \\
  t &=&\frac{1}{D}\exp \left( D\tau \right) \coth \left( \alpha _{2}+%
\sqrt{2}y\right)   ,  \label{coordt3}
\end{eqnarray}%
where we have denoted $B=\coth \alpha _{2}$. The horizon coordinates obey $%
t^{2}\rho ^{2}>1$. The GM space-time in this case also covers only partially
the horizon space-time (\ref{Horizons_sol}). In this part of the horizon
space-time $K_{\mathbf{8}}$ is time-like, as well as $K_{7\mathbf{,9}}$.

(b3) For $\varepsilon =-1$ and $B^{2}<1$ (implying $\mathit{{sgn}\left(
D^{2}\right) =-1}$) the horizon coordinates $\left( \rho ,~t\right) $ are
related to the GM coordinates by a complex transformation: 
\begin{eqnarray}
\rho &=&-iD\exp \left( -D\tau \right) \cosh \left( \alpha _{3}+\sqrt{2}%
y\right)   ,  \nonumber \\
t &=&\frac{1}{D}\exp \left( D\tau \right) \tanh \left( \alpha _{3}+\sqrt{2}%
y\right)   ,  \label{coordt2}
\end{eqnarray}%
with $B=\tanh $ $\alpha _{3}$. Note that in this case $D$ is purely
imaginary, which implies $t^{2}\rho ^{2}<0\,$. The coordinate transformation
being complex, this case is the closest analogue of the general relativistic
result that the Bertotti-Robinson metric is related to the horizon region of
the extremal Reissner-Nordstr\"{o}m metric (see the striking similarity with
the structure of Eqs. (\ref{trafo}) in Appendix C).

Remembering that the transformation (\ref{coorda}) classified as (a) relates
the degenerated horizon metric of the 5d black hole (\ref{Horizons_sol})
with $\varepsilon =-1$ to the GM space-time with $\varepsilon =-1$ and $%
B=\pm 1$, we see that all possible cases of the GM metric with cosmological
constant ($\varepsilon =\pm 1,~B$ arbitrary) are covered by our analysis.
All GM metrics with $\varepsilon =-1$ are covered in (a), (b2), (b3), while
the GM metrics with $\varepsilon =1$ in (b1).

In the cases with $\varepsilon =-1$ the static character of the 5d black
hole metric both above and below the degenerated horizons is assured by $K_{%
\mathbf{7}}$, which also remains a time-like Killing vector for the horizon
metric. In all these cases $K_{\mathbf{9}}$ is also time-like, while $K_{%
\mathbf{8}}$ is time-like for (b2), space-like for (b3), and its causal
character depends on the actual value of the coordinates for (a), as for
this transformation (\ref{coorda}) $\rho ^{2}t^{2}>0$.

As remarked earlier, for $\varepsilon =1$ [case b1)] the coordinate
transformation (\ref{coordt1}) relates the static GM metric to the static
region of the horizon metric.

In each case the GM metric can be related to the horizon metric either by a
real [cases (a), (b1), (b2)], or by a complex coordinate transformation
[case (b3)]. The latter case is similar to the general relativistic analogy
between the Bertotti-Robinson space-time and the degenerated horizon region
of the extremal Reissner-Nordstr\"{o}m black hole shown by a complex
transformation (see Appendix C). Both there and in our case (b3) this is
understood in the following sense: although the emerging coordinates are
complex, only their real subset is considered in the line-element.

\section{Concluding Remarks}

We have shown that the 5d GM space-time, which contains the Einstein brane
as boundary, although violates the 5d Birkhoff theorem (being an 5d vacuum
space-time different from (\ref{Sch}) and with an embedded \textit{static}
Friedmann brane), obeys the theorem in the following weaker sense. For all
cases of the GM space-time parameters a specific 5d black hole metric can be
found for which the GM metric is related to its degenerated horizon region
either by a real or a complex coordinate transformation. We have proven this
result by explicitly constructing the respective coordinate transformations.

For a positive 5d cosmological constant ($\varepsilon =1$), the GM metric
represents the static region of the horizon metric which approximates the 5d
SchwarzSchild- de Sitter 5d black hole degenerate horizon region.

A negative 5d cosmological constant ($\varepsilon =-1$) is far more
acceptable from a brane point of view as it gives a small cosmological
constant $\Lambda $ on the brane through the relation:

\begin{equation}
2\Lambda =\kappa ^{2}\lambda +3\varepsilon \Gamma ^{2}~,
\end{equation}%
(where $\lambda =6\kappa ^{2}/\widetilde{\kappa }^{4}$ is the brane tension,
known to have a high value \cite{branetension}, and $\kappa ^{2}$ is the
brane gravitational constraint). For this case we have shown that the GM
metric is related to the static horizon metric representing the region close
to the degenerated horizon of a 5d SchwarzSchild - anti de Sitter -like
black hole with curvature index $k=-1$.

The generic result established in this paper according to which the GM
space-time containing the Einstein brane is the degenerated horizon region
of the SchwarzSchild - (anti) de Sitter 5d black hole, is in close analogy
with the general relativistic result, that the Bertotti-Robinson space-time
generated by a covariantly constant electromagnetic field is the degenerated
horizon region of the extremal Reissner-Nordstr\"{o}m space-time.

Our result re-establishes the validity of the Birkhoff theorem in 5d,
although in a weaker sense.

\ack 
This work was supported by OTKA grants no. 46939 and 69036. L\'{A}G was
further supported by the J\'{a}nos Bolyai Grant of the Hungarian Academy of
Sciences.

\appendix

\section{Horizons in the space-time (\protect\ref{Sch})}

The metrics with constant curvature (\ref{Sch}) admitting branes with
constants spatial curvature for the various possible values of $\varepsilon $
and $k$ and sign of the mass parameter $m$ in certain cases describe black
holes with horizons given in the Tables \ref{table1}-\ref{table3}. 
\begin{table}[h]
\caption{The location (given by the $r$ coordinate) of the horizons with
vanishing 5d cosmological constant ($\protect\varepsilon =0$). }
\label{table1}
\begin{center}
\begin{tabular}{c|ccc}
& $m<0$ & $m=0$ & $m>0$ \\ \hline
&  &  &  \\ 
$k=-1$ & $\sqrt{-2m}$ & flat metric & $-$ \\ 
$k=0$ & $-$ & ill-defined metric & $-$ \\ 
$k=1$ & $-$ & flat metric & $\sqrt{2m}$%
\end{tabular}%
\end{center}
\end{table}
\begin{table}[h]
\caption{The location (given by $\Gamma r$) of the horizons with positive 5d
cosmological constant ($\protect\varepsilon =1$). }
\begin{center}
\begin{tabular}{c|ccc}
& $m<0$ & $m=0$ & $m>0$ \\ \hline
&  &  &  \\ 
$k=-1$ & $\sqrt{-1+\sqrt{1-4m\Gamma ^{2}}}$ & $-$ & $-$ \\ 
$k=0$ & $\sqrt[4]{-4m\Gamma ^{2}}$ & $-$ & $-$ \\ 
$k=1$ & $\sqrt{1+\sqrt{1-4m\Gamma ^{2}}}$ & $\sqrt{2}$ & $\sqrt{1\pm \sqrt{%
1-4m\Gamma ^{2}}}$%
\end{tabular}%
\end{center}
\label{table2}
\end{table}
\begin{table}[h]
\caption{Same as in the Table \protect\ref{table2} with $\protect\varepsilon %
=-1$.}
\begin{center}
\begin{tabular}{c|ccc}
& $m<0$ & $m=0$ & $m>0$ \\ \hline
&  &  &  \\ 
$k=-1$ & $\sqrt{1\pm \sqrt{1+4m\Gamma ^{2}}}$ & $\sqrt{2}$ & $\sqrt{1+\sqrt{%
1+4m\Gamma ^{2}}}$ \\ 
$k=0$ & $-$ & $-$ & $\sqrt[4]{4m\Gamma ^{2}}$ \\ 
$k=1$ & $-$ & $-$ & $\sqrt{-1+\sqrt{1+4m\Gamma ^{2}}}$%
\end{tabular}%
\end{center}
\label{table3}
\end{table}

\section{The Killing algebra of the horizon metric}

The solution of the Killing equation gives the following independent Killing
vectors for the horizon metric (\ref{Horizons_sol}), written in the
coordinate basis $\left( t,~\rho ,~\chi ,~\theta ,~\varphi \right) $: 
\label{KillingVectors}
\begin{eqnarray}
K_{\mathbf{1}} &=&\left( 0,0,0,0,1\right)   , \\
K_{\mathbf{2}} &=&\left( 0,0,0,-\cos \varphi ,\cot \theta \sin \varphi
\right)   , \\
K_{\mathbf{3}} &=&\left( 0,0,0,\sin \varphi ,\cot \theta \cos \varphi
\right)   , \\
K_{\mathbf{4}} &=&\left( 0,0,-\cos \theta ,\sin \theta   \partial
_{\chi }\ln \mathcal{H},0\right)   , \\
K_{\mathbf{5}} &=&\left( 0,0,\sin \theta \sin \varphi ,\cos \theta \sin
\varphi   \partial _{\chi }\ln \mathcal{H},\frac{\cos \varphi }{\sin
\theta }  \partial _{\chi }\ln \mathcal{H}\right)   , \\
K_{\mathbf{6}} &=&\left( 0,0,\sin \theta \cos \varphi ,\cos \theta \cos
\varphi   \partial _{\chi }\ln \mathcal{H},-\frac{\sin \varphi }{\sin
\theta }  \partial _{\chi }\ln \mathcal{H}\right)   , \\
K_{\mathbf{7}} &=&\left( 1  ,0,0,0,0\right)   , \\
K_{\mathbf{8}} &=&\left( t,-\rho ,0,0,0\right)   , \\
K_{\mathbf{9}} &=&\left( \frac{t^{2}}{2}+\frac{1}{2\rho ^{2}}  ,-t\rho
,0,0,0\right)   .
\end{eqnarray}%
The Killing vectors $K_{\mathbf{1-6}}$ are the usual cosmological symmetries
(representing rotations and quasi-translations), and they are space-like. In
order to find out the causal character of the rest of the Killing vectors we
calculate their length in the horizon metric: 
\begin{eqnarray}
g\left( K_{\mathbf{7}},K_{\mathbf{7}}\right) &=&\frac{\varepsilon }{2\Gamma
^{2}}\rho ^{2}  , \\
g\left( K_{\mathbf{8}},K_{\mathbf{8}}\right) &=&\frac{\varepsilon }{2\Gamma
^{2}}\left( \rho ^{2}t^{2}-1\right)   , \\
g\left( K_{\mathbf{9}},K_{\mathbf{9}}\right) &=&\frac{\varepsilon }{8\Gamma
^{2}\rho ^{2}}\left( \rho ^{2}t^{2}-1\right) ^{2}  .
\end{eqnarray}%
It is obvious that $K_{\mathbf{7}}$ and $K_{\mathbf{9}}$ are time-like for $%
\varepsilon =-1$ and space-like for $\varepsilon =1$, while the causal
character of $K_{\mathbf{8}}$ depends on the sign of the product $%
\varepsilon \left( \rho ^{2}t^{2}-1\right) $. The horizon metric is static
in all cases excepting when $\varepsilon =1$ and $\rho ^{2}t^{2}>1$. Also
the locus $\rho t=\pm 1$ is a Killing horizon for $K_{\mathbf{8}}$.

The Killing vectors $K_{\mathbf{1-7}}$ are also Killing vectors for the
black hole metric (\ref{Sch}). While $K_{\mathbf{1-6}}$ remain space-like,
the causal character of $K_{\mathbf{7}}$ depends on the region of
space-time: calculated with the metric (\ref{Sch}) $g\left( K_{\mathbf{7}%
},K_{\mathbf{7}}\right) =-f$. Therefore $K_{\mathbf{7}}$ is time-like if
there is no horizon; time-like above the horizon and space-like below, if
there is one horizon; and time-like above the exterior horizon and below the
inner horizon, space-like between the two horizons, when there are two
horizons (it is time-like everywhere excepting the horizon for degenerated
horizons); finally on any horizon is null, thus the event horizons are also
Killing horizons for $K_{\mathbf{7}}$ in the black hole metrics (\ref{Sch}),
a property which is lost in the approximate horizon metric.

Having the additional $K_{\mathbf{8,9}}$ Killing vectors, the horizon metric
has more symmetries, than the full black hole metric.

The Killing algebra is given by

\begin{eqnarray}
\left[ K_{\mathbf{i}},K_{\mathbf{j}}\right] &=&\varepsilon _{ijk}K_{\mathbf{k%
}}  , \\
\left[ K_{\mathbf{{3}+{i}}},K_{\mathbf{{3}+{j}}}\right] &=&\varepsilon  {
}\varepsilon _{ijk}K_{\mathbf{k}}  , \\
\left[ K_{\mathbf{i}},K_{\mathbf{{3}+{j}}}\right] &=&\varepsilon _{ijk}K_{%
\mathbf{{3}+{k}}}  , \\
\left[ K_{\mathbf{{6}+{i}}},K_{\mathbf{j}}\right] &=&0=\left[ K_{\mathbf{{6}+%
{i}}},K_{\mathbf{{3}+{j}}}\right]   , \\
\left[ K_{\mathbf{7}},K_{\mathbf{8}}\right] &=&K_{\mathbf{7}}  , \\
\left[ K_{\mathbf{8}},K_{\mathbf{9}}\right] &=&K_{\mathbf{9}}  , \\
\left[ K_{\mathbf{7}},K_{\mathbf{9}}\right] &=&K_{\mathbf{8}}  ,
\end{eqnarray}%
and is classified in Table \ref{table4}. 
\begin{table}[h]
\caption{Killing algebras of the black hole metric (upper row) and horizon
metric (bottom row) for $\protect\varepsilon =\pm 1$.}
\label{table4}
\begin{center}
\begin{tabular}{c|cc}
$\varepsilon $ & $1$ & $-1$ \\ \hline
$K_\mathbf{{{1}-{7}}}$ & $so\left( 4\right) \oplus $ & $so\left( 1,3\right)
\oplus $ \\ 
$K_\mathbf{{{1}-{9}}}$ & $so\left( 4\right) \oplus so\left( 1,2\right) $ & $%
so\left( 1,3\right) \oplus so\left( 1,2\right) $%
\end{tabular}%
\end{center}
\end{table}

The Killing vectors $K_{\mathbf{1-6}}$ of the horizon metric and $K_{\mathbf{%
1-6}}^{GM}$ of the GM metric are identical and the Killing vectors $K_{%
\mathbf{7-9}}$ are also related to the Killing vectors $K_{\mathbf{7-9}%
}^{GM} $ (specified for $A=1$). In order to establish these relations, $K_{%
\mathbf{7-9}}$ have to be transformed in the GM coordinate basis by the
coordinate transformations derived in Sections 3 and 4:

(a) Applying the coordinate transformation (\ref{coorda}) on $K_{\mathbf{7-9}%
}$ we find 
\label{KillingVectorsrel}
\begin{equation}
K_{\mathbf{7}}^{GM}=K_{\mathbf{7}}  ,\quad K_{\mathbf{8}}^{GM}=-K_{%
\mathbf{9}}  ,\quad K_{\mathbf{9}}^{GM}=-K_{\mathbf{8}}  .
\end{equation}

(b1) Applying (\ref{coordt1}): 
\label{KillingVectorsepsilon1}
\begin{eqnarray}
K_{\mathbf{7}}^{GM} &=&K_{\mathbf{8}}  , \\
K_{\mathbf{8,9}}^{GM} &=&DK_{\mathbf{9}}\pm \frac{1}{2D}K_{\mathbf{7}}{ 
}.
\end{eqnarray}

(b2) Applying (\ref{coordt3}): 
\begin{eqnarray}
K_{\mathbf{7}}^{GM} &=&K_{\mathbf{8}}  , \\
K_{\mathbf{8,9}}^{GM} &=&-DK_{\mathbf{9}}\pm \frac{1}{2D}K_{\mathbf{7}}{
}.
\end{eqnarray}

(b3) Finally applying (\ref{coordt2}):%
\begin{eqnarray}
K_{\mathbf{7}}^{GM} &=&iK_{\mathbf{8}}  , \\
K_{\mathbf{8}}^{GM} &=&DK_{\mathbf{9}}-\frac{1}{2D}K_{\mathbf{7}}  , \\
K_{\mathbf{9}}^{GM} &=&i\left( DK_{\mathbf{9}}+\frac{1}{2D}K_{\mathbf{7}%
}\right)   .
\end{eqnarray}%
Thus in the cases (a), (b1), (b2) $K_{\mathbf{7-9}}^{GM}$ are linear
combinations with constant real coefficients of $K_{\mathbf{7-9}}$, while in
the case (b3) the linear combination is complex.

\section{General relativistic analogy: the Bertotti-Robinson metric as the
horizon region of the extremal Reissner-Nordstr\"{o}m space-time}

The Reissner-Nordstr\"{o}m metric describes the spherically symmetric,
static electro-vacuum exterior of a point mass $m$ with electric charge $q$.
The two horizons degenerate into a single one located at $r=m$ in the
extremal case, when $q=m\,$. Then the line element takes the form%
\begin{equation}
ds_{RN}^{2}=-\left( 1-\frac{m}{r}\right) ^{2}dt^{2}+\left( 1-\frac{m}{r}%
\right) ^{-2}dr^{2}+r^{2}d\Omega ^{2}  ,  \label{RN}
\end{equation}%
with $d\Omega ^{2}$ the infinitesimal solid angle on the unit $2$-sphere. In
order to approximate the metric (\ref{RN}) in the vicinity of the horizon,
it is useful to introduce a new coordinate $\rho =r-m$ \cite{Brill}, in
terms of which the line element (\ref{RN}) becomes%
\begin{equation}
ds_{RN}^{2}=-\left( \frac{\rho }{\rho +m}\right) ^{2}dt^{2}+\left( \frac{%
\rho }{\rho +m}\right) ^{-2}d\rho ^{2}+\left( \rho +m\right) ^{2}d\Omega ^{2}%
  .
\end{equation}%
Close to the horizon ($\rho \approx 0$) the extremal Reissner-Nordstr\"{o}m
space-time is approximated as:%
\begin{equation}
ds_{hRN}^{2}=-\left( \frac{\rho }{m}\right) ^{2}dt^{2}+\left( \frac{\rho }{m}%
\right) ^{-2}d\rho ^{2}+m^{2}d\Omega ^{2}  .  \label{hRN}
\end{equation}%
The sequence of transformations i.) $t^{^{\prime }}=it$; ii.) $\rho =m\exp
\left( -\tau ^{\prime }\right) \cosh z$, $t^{\prime }=m\exp \left( \tau
^{\prime }\right) \tanh z$; iii.) $\tau =i\tau ^{\prime }$ brings the metric
into the form \cite{Brill}: 
\begin{equation}
ds_{hRN}^{2}=m^{2}\left[ -\cosh ^{2}z~d\tau ^{2}+dz^{2}+d\Omega ^{2}\right] 
  .  \label{BRspec}
\end{equation}%
The sequence of coordinate transformations can also be given as%
\begin{eqnarray}
\rho &=&m\exp \left( i\tau \right) \cosh z~,  \nonumber \\
t &=&-im\exp \left( -i\tau \right) \tanh z  ,  \label{trafo}
\end{eqnarray}%
with the inverse:%
\begin{equation}
z=arcsinh\left( \frac{i\rho t}{m^{2}}\right) ,{\qquad }2i\tau =\ln 
\frac{m^{2}\rho ^{2}}{m^{4}-t^{2}\rho ^{2}}  .  \label{trafoinv}
\end{equation}

The energy-momentum tensor of the extremal Reissner-Nordstr\"{o}m space-time
in the $\left( t,r,\theta ,\varphi \right) $ coordinate system is%
\begin{equation}
T_{b}^{a}=\frac{m^{2}}{r^{4}}\mathit{{diag}\left( -1,-1,1,1\right)   .}
\end{equation}%
With the coordinate $\rho =r-m$ and for $\rho \approx 0$ this approximates as%
\begin{equation}
T_{b}^{a}=\frac{1}{m^{2}}\mathit{{diag}\left( -1,-1,1,1\right)   .}
\label{ThRN}
\end{equation}%
The approximate horizon metric (\ref{hRN}) solves the Einstein equations for
the above energy-momentum tensor. Moreover, after applying the complex
coordinate transformation (\ref{trafo}) the energy-momentum tensor will have
the same form (\ref{ThRN}).

This describes a pure electric field, as required. In order to see this, we
note that in the coordinates $\left( t,r,\theta ,\varphi \right) $ the only
non-vanishing components of the Maxwell tensor for the Reissner-Nordstr\"{o}%
m space-time are%
\begin{equation}
F_{tr}=-F_{rt}=-\frac{q}{r^{2}}  .
\end{equation}%
In the extremal case $q=m$ and the degenerated horizon is at $r=m$, such
that the non-vanishing components of the Maxwell tensor become%
\begin{equation}
F_{tr}=-F_{rt}=-\frac{1}{m}  .  \label{horizonE}
\end{equation}

The Bertotti-Robinson space-time \cite{Bertotti}, \cite{Robinson} represents
the product of two Riemannian $2$-surfaces with constant curvature radius,
generated by a covariantly constant electromagnetic field in the presence of
a cosmological constant. Its generic form is given by the line-element%
\begin{equation}
ds_{BR}^{2}=-\left( 1+\frac{x^{2}}{r_{+}^{2}}\right) dt^{2}+\left( 1+\frac{%
x^{2}}{r_{+}^{2}}\right) ^{-1}dx^{2}+r_{-}^{2}d\Omega ^{2}~.  \label{BR}
\end{equation}%
By performing the coordinate transformations%
\begin{equation}
arcsinh\frac{x}{r_{+}}=z~,\qquad t=r_{+}\tau
\end{equation}%
which gives $1+x^{2}/r_{+}^{2}=\cosh ^{2}z$, the Bertotti-Robinson metric
becomes%
\begin{equation}
ds_{BR}^{2}=r_{+}^{2}~[-\cosh ^{2}z~d\tau ^{2}+dz^{2}]+r_{-}^{2}~d\Omega
^{2}~,
\end{equation}%
which agrees with the metric (\ref{BRspec}) in the case when the two
Riemannian surfaces have the same curvature radii: 
\begin{equation}
r_{+}=r_{-}=m~.
\end{equation}%
The equality of the two curvature radii is equivalent with the vanishing of
the cosmological constant in the Bertotti-Robinson solution, which therefore
represents the space-time generated by a pure electromagnetic field.

The Maxwell tensor for the Bertotti-Robinson space-time is given by Eq. (17)
in \cite{Bertotti}, representing parallel electric and magnetic fields. In
the $\left( \tau ,z,\theta ,\varphi \right) $ coordinate system the
energy-momentum tensor is given by 
\begin{equation}
T_{b}^{a}=\mu \mathit{{diag}\left( -1,-1,1,1\right)   .}
\end{equation}%
Here $\mu $ $=1/m^{2}$ [see (\ref{ThRN})] is related to the two invariants
of the electromagnetic field as%
\begin{equation}
\mu ^{2}=\left( \mathbf{{h}^{2}-{e}^{2}}\right) ^{2}+\left( 2\mathbf{eh}%
\right) ^{2}  .
\end{equation}%
Since the energy-momentum tensor only depends on $\rho $, this is what the
geometry determinates. Thus another key information about the
electromagnetic field, represented by the parameter%
\begin{equation}
\alpha =-\frac{1}{2}\arctan \frac{2\mathbf{eh}}{\mathbf{{h}^{2}-{e}^{2}}}%
  ,
\end{equation}%
remains undeterminated. With the convenient choice of $\alpha $ the
electromagnetic field can be chosen as a pure electric field with $\mathbf{{e%
}^{2}=}1/m^{2}$ in perfect agreement with the source (\ref{horizonE}) of the
horizon metric.

\end{document}